\documentclass[10pt]{article}
\usepackage{graphicx}
\usepackage{cite}
\usepackage{epstopdf}
\newcommand{\pt}{p_{\rm T}}
\newcommand{\kt}{k_{\rm T}}

\def\Title#1{\begin{center} {\Large #1 } \end{center}}
\def\Author#1{\begin{center}{ \sc #1} \end{center}}
\def\Address#1{\begin{center}{ \it #1} \end{center}}

\newenvironment{Abstract}{\begin{quotation} \begin{center} 
             \large ABSTRACT \end{center}\bigskip 
      \begin{center}\begin{large}}{\end{large}\end{center} \end{quotation}}

\newenvironment{Presented}{\begin{quotation} \begin{center} 
             PRESENTED AT\end{center}\bigskip 
      \begin{center}\begin{large}}{\end{large}\end{center} \end{quotation}}

\def\beq{\begin{equation}}
\def\eeq#1{\label{#1}\end{equation}}
\def\eeqn{\end{equation}}

\def\beqa{\begin{eqnarray}}
\def\eeqa#1{\label{#1}\end{eqnarray}}
\def\eeqan{\end{eqnarray}}

\let\bar=\overbar

\def\Dslash{\not{\hbox{\kern-4pt $D$}}}
\def\dslash{\not{\hbox{\kern-2pt $\del$}}}

\def\msb{{\bar{\ssstyle M \kern -1pt S}}}

\usepackage{color}

\def\affiliation{
On behalf of the ALICE Collaboration, \\
INFN Torino \\
Via P. Giuria 1, 10125 Torino, ITALY }

\begin{document}

\large
\begin{titlepage}

\vfill
\Title{ Heavy-flavour production and multiplicity dependence in pp and p--Pb collisions with ALICE.  }
\vfill

\Author{ Elena Bruna  }
\Address{\affiliation}
\vfill
\begin{Abstract}

The production of heavy quarks in pp collisions provides a precision test of perturbative QCD calculations at the LHC energies. More complex collision systems like p--Pb collisions allow studies of cold nuclear matter effects, such as modifications of the parton distribution functions at small x and of the $\kt$ broadening effect. 

We present the ALICE results of prompt D-meson production as a function of the charged-particle multiplicity, in pp and p--Pb collisions at $\sqrt{s}=7$ TeV and $\sqrt{s_{NN}}=5.02$ TeV respectively. The per-event yield of D mesons in different multiplicity and $\pt$ intervals are compared for pp and p--Pb collisions to study the contribution of multi-parton interactions to open-charm production. 
Angular correlations of prompt D mesons and heavy-flavour decay electrons with charged hadrons in pp and p--Pb collisions are also shown in different kinematic ranges and compared to pQCD models. These measurements provide information on the charm fragmentation processes, on cold nuclear matter effects on charm production, and on the participation of charm in the collective motion arising in small collision systems like p--Pb.

\end{Abstract}
\vfill

\begin{Presented}
The Second Annual Conference\\
 on Large Hadron Collider Physics \\
Columbia University, New York, U.S.A \\ 
June 2-7, 2014
\end{Presented}
\vfill
\end{titlepage}
\def\thefootnote{\fnsymbol{footnote}}
\setcounter{footnote}{0}
%

\normalsize 


\section{Introduction}
 Heavy quarks such as charm and beauty are produced in hard parton scattering processes in the initial stages of hadronic collisions. Their production in pp collisions at the LHC provides, therefore, a precision test of perturbative QCD calculations at the highest collision energies. 
 More complex collision systems like p--Pb collisions make it possible to study cold nuclear matter and collective effects as well as to characterize the nuclear parton distribution functions at small x in the region where gluons are expected to saturate, as well as to gain insight into the $\kt$ broadening due to soft scatterings of the partons in the incoming nucleus. 
  Heavy-quark production in pp and p--Pb collisions can also help understanding the role of  multiple initial hard parton scatterings. In particular, studies of heavy-quark production as a function of charged-particle multiplicity in pp and p--Pb collisions are expected to be sensitive to the interplay between hard and soft QCD processes and therefore to the effects of multi-parton interactions affecting the hard momentum scale relevant for heavy-quark production.
In addition, measurements of angular correlations of D mesons or heavy-flavour decay electrons with charged hadrons are interesting in both pp and p--Pb collisions at the LHC. D-hadron correlations provide information on the heavy-quark fragmentation, while electron-hadron correlations can also be used as a tool to measure the relative charm and beauty contributions to the heavy-flavour electron yield in pp collisions~\cite{electronB2.76}. Such measurements also provide information on the participation of charm in the collective motion arising in small collision systems like p--Pb.

\section{Heavy quark production in pp and p--Pb collisions at the LHC}
Heavy flavours are measured in ALICE via exclusive and semi-inclusive reconstruction techniques.
D mesons are detected via their hadronic decays with an analysis based on displaced secondary vertices and particle identification. The correction for D mesons coming from feed-down of beauty hadrons is based on FONLL calculations~\cite{fonll}. Electrons and muons from heavy-flavour hadron decays (charm and beauty) are detected in ALICE at mid- and forward rapidity respectively. 
The background contribution in the muon inclusive spectrum, dominated by
muons from light hadron decays, is estimated via a data-driven method using
as input the $\pi^\pm$ and K$^\pm$  yields measured at mid-rapidity with ALICE. In the
inclusive electron spectrum, the main sources of background are photon
conversions and $\pi^0$ Dalitz decays. They are statistically subtracted by using
two alternative methods: the cocktail (a calculation of the non-heavy-flavour
electron sources based on measured $\pi^0$ spectra) and the invariant-mass
approach (based on the measurement of electrons from photon conversions and Dalitz decays via
low-mass e$^+$e$^-$ pairs).
Moreover, selections based on displaced track impact parameters are also applied to measure electrons coming from decays of beauty hadrons. 

\begin{figure}[htb]
\centering
\includegraphics[angle=0, width=6.cm]{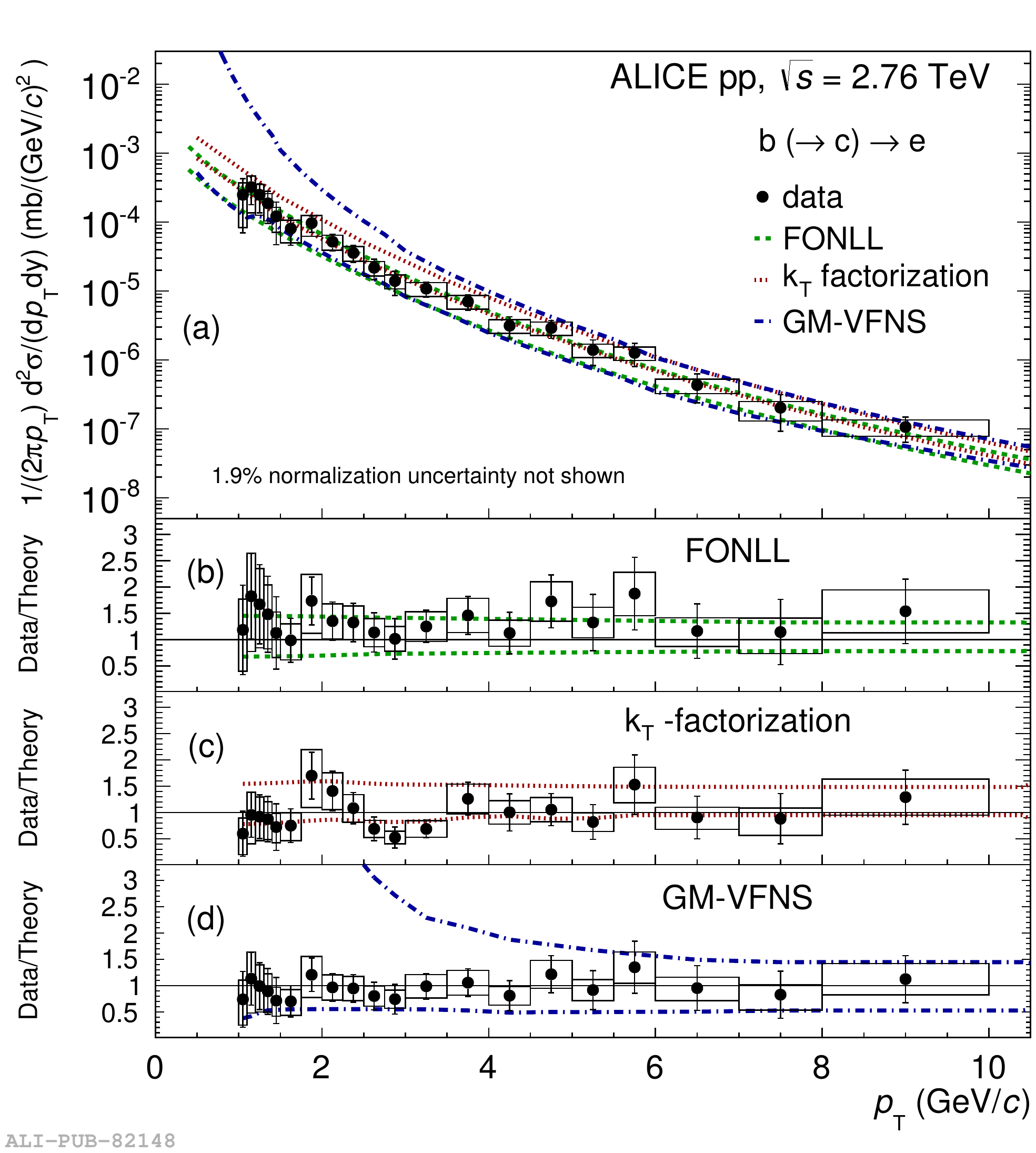}
\includegraphics[angle=0, width=7.5cm]{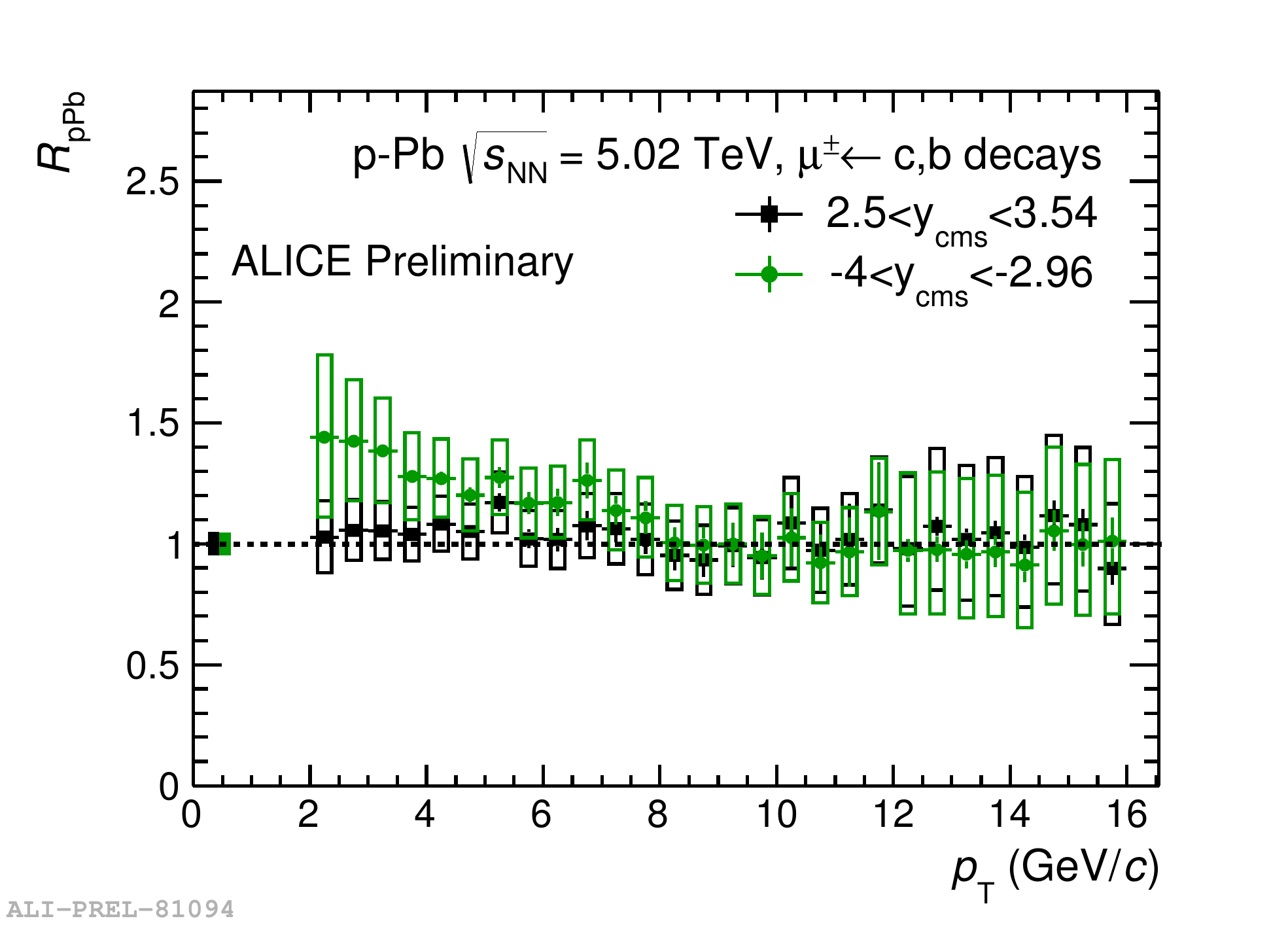}
\caption{Left: cross section of electrons from beauty-hadron decays as a function of $\pt$ in pp collisions at $\sqrt s$=2.76 TeV compared to different pQCD calculations~\cite{fonll, GM-VFNS, kt}. Right: $R_{\rm {pPb}}$ of muons from heavy-flavour decays as a function of $\pt$ at both forward (black) and backward (green) rapidity. }
\label{fig:BeleMu}
\end{figure}

Cross section measurements of prompt D mesons and electrons and muons from decays of heavy-flavoured particles were carried out with ALICE in pp collisions at $\sqrt s$=7 and 2.76 TeV~\cite{electronB 2.76,electronB,electronHF,Ds,JpsiB,D2.76,muonHF,D7} and are well described by different perturbative QCD  calculations at both energies~\cite{fonll, GM-VFNS, kt}. An example is reported in Fig.~\ref{fig:BeleMu} (left), where the $\pt$-differential cross section of electrons from beauty-hadron decays measured at mid-rapidity is shown for pp collisions at $\sqrt s$=2.76 TeV and compared to different calculations based on perturbative QCD.

It is now interesting to study the heavy-flavour production in p--Pb collisions at $\sqrt s_{NN}$=5.02 TeV, where the modified parton distribution functions due to the presence of the nucleus are expected to affect the heavy-quark yield and $\pt$ distribution in p--Pb relative to pp collisions. In particular, by measuring heavy quarks in different rapidity ranges it is possible to access different x  regimes. This can be observed in the right panel of Fig.~\ref{fig:BeleMu}, where the nuclear modification factor $R_{\rm {pPb}}$ of muons from heavy-flavour hadron decays, defined as the ratio of the muon cross section in p--Pb collisions to the same cross section in pp scaled by the Pb-nucleus mass number A, is reported as a function of $\pt$ at both forward (black) and backward (green) rapidities. The $R_{\rm {pPb}}$ measurements at different rapidities allow studies of initial-state effects relevant at low $\pt$ ($\pt<4$ GeV/$c$) due to different x ranges explored depending on the rapidity range (x$\sim$10$^{-5}$ and $\sim$10$^{-2}$ for forward and backward regions respectively). The $R_{\rm {pPb}}$ of heavy-flavour decay muons and prompt D mesons is described within uncertainties by different models of initial state effects (see~\cite{RpPbD} and references therein).

\section{Heavy-flavour production as a function of charged-particle multiplicity}

\begin{figure}[htb]
\centering
\includegraphics[angle=0, width=6.5cm]{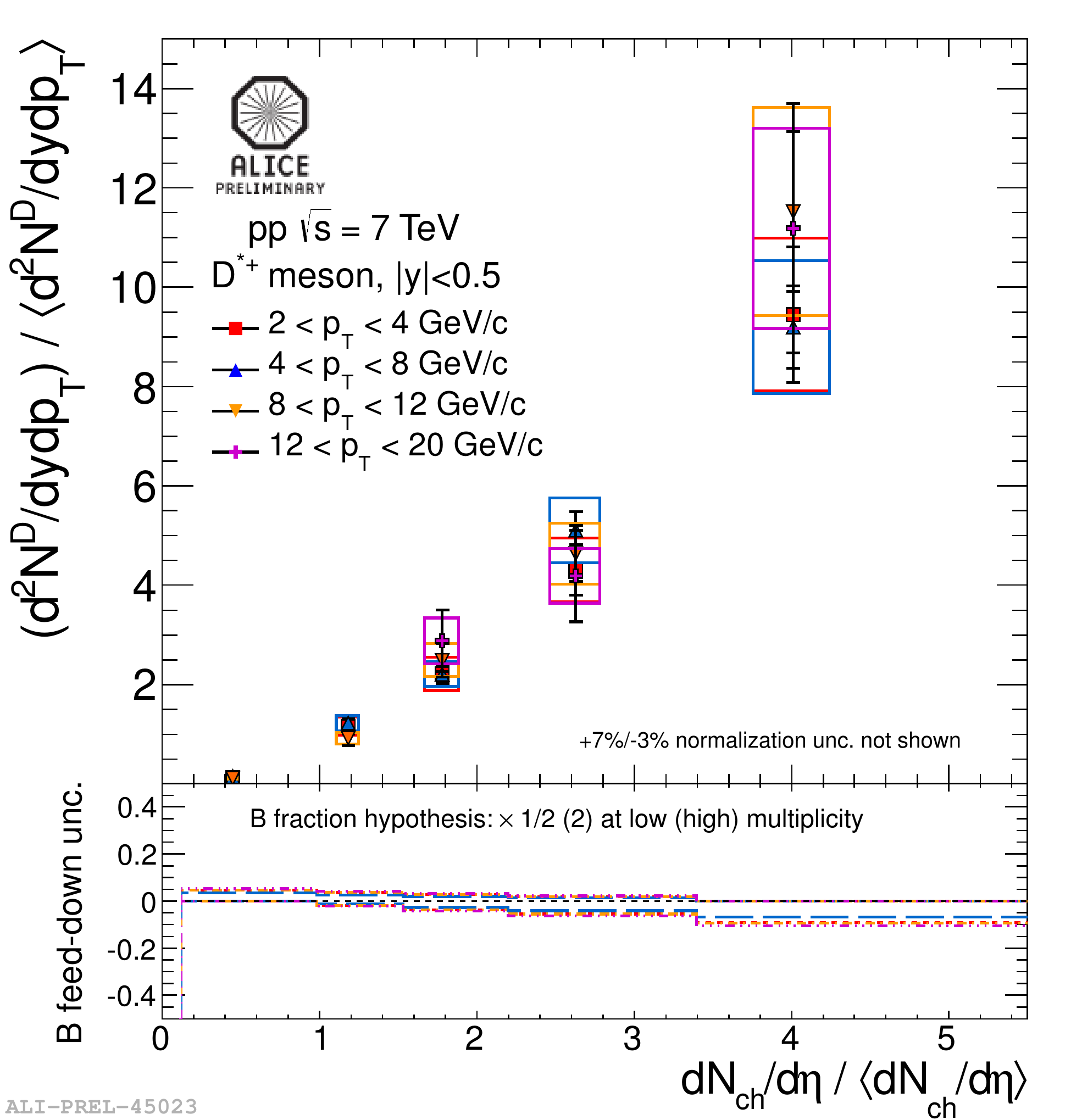}
\includegraphics[angle=0, width=6.5cm]{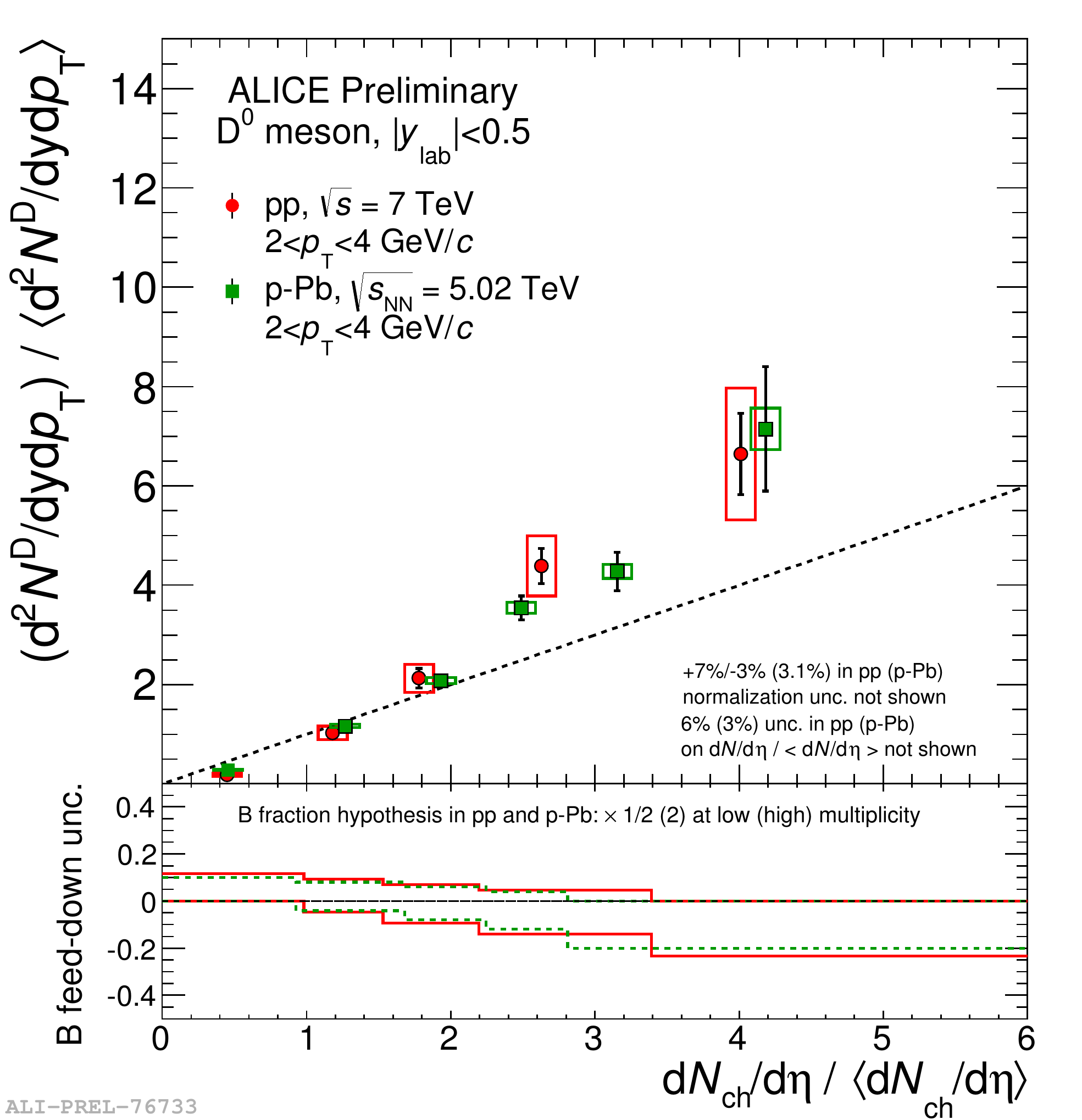}
\caption{Left:  self-normalized D$^{*+}$ yield as a function of multiplicity in pp collisions for different $\pt$ ranges. Right: self-normalized D$^{0}$ yield as a function of multiplicity in pp (red) and p--Pb (green) collisions for D$^{0}$ mesons with $2<\pt<4$ GeV/$c$.}
\label{fig:mult}
\end{figure}

The question we now want to ask is whether the production of heavy quarks is affected by the presence of additional hard scatterings occurring between incoming partons in the same hadronic collision (Multi-Parton Interactions, MPIs).
It has been observed that MPIs play a relevant role in particle production at LHC energies. In particular, CMS results on jets and underlying events showed a better agreement with models including MPIs~\cite{cmsMPI} and measurements with ALICE of minijets indicate an increase of MPIs with increasing multiplicity of charged particles in a pp event~\cite{minijetALICE}.
Studies of effects of MPIs on heavy-flavour production were reported at the SPS energies by the NA27 Collaboration, which showed that events with charmed hadrons have on average larger charged-particle multiplicity~\cite{NA27}, suggesting the presence of additional parton scatterings occurring on the hard scale relevant for heavy-quark production. At the LHC energies, the LHCb Collaboration observed that the double charm production agrees better with models including double parton scatterings~\cite{LHCbMPI} in pp collisions. Effects of MPIs on quarkonium production were observed by the ALICE Collaboration which measured an increase of the J/$\psi$  yield as a function of event multiplicity in pp collisions~\cite{ALICEjpsimult}, and by the CMS Collaboration, which reported an increase of the yield of bottomonium states with increasing event multiplicity in both pp and p--Pb collisions~\cite{CMSUpsilonimult}. 
Similar studies as a function of the event multiplicity were recently performed by the ALICE Collaboration for prompt D mesons (D$^0$, D$^+$, D$^{*+}$) in both pp and p--Pb collisions. In Fig.~\ref{fig:mult} (left), the D$^{*+}$-meson per-event yield in multiplicity intervals normalized to the same yield obtained in the multiplicity-integrated sample is reported as a function of the charged-particle multiplicity normalized by its average measured in multiplicity-integrated events. The results in the different $\pt$ ranges shown in the figure are in agreement within the uncertainties. An increasing trend of the D-meson yield as a function of charged-particle multiplicity in pp collisions is observed, suggesting that MPIs which substantially contribute to high-multiplicity events are affecting heavy-flavour production. A similar trend is observed in p--Pb collisions, as shown in the right panel of Fig.~\ref{fig:mult} for the D$^0$ meson, although in this case a higher number of binary nucleon-nucleon collisions is expected to contribute to high-multiplicity events.
The observed trend in pp collisions is also similar to that observed for the J/$\psi$ yield as a function of multiplicity~\cite{ALICEjpsimult}.

\section{Heavy-flavour azimuthal correlations}
\begin{figure}[htb]
\centering
\includegraphics[angle=0, width=6.5cm]{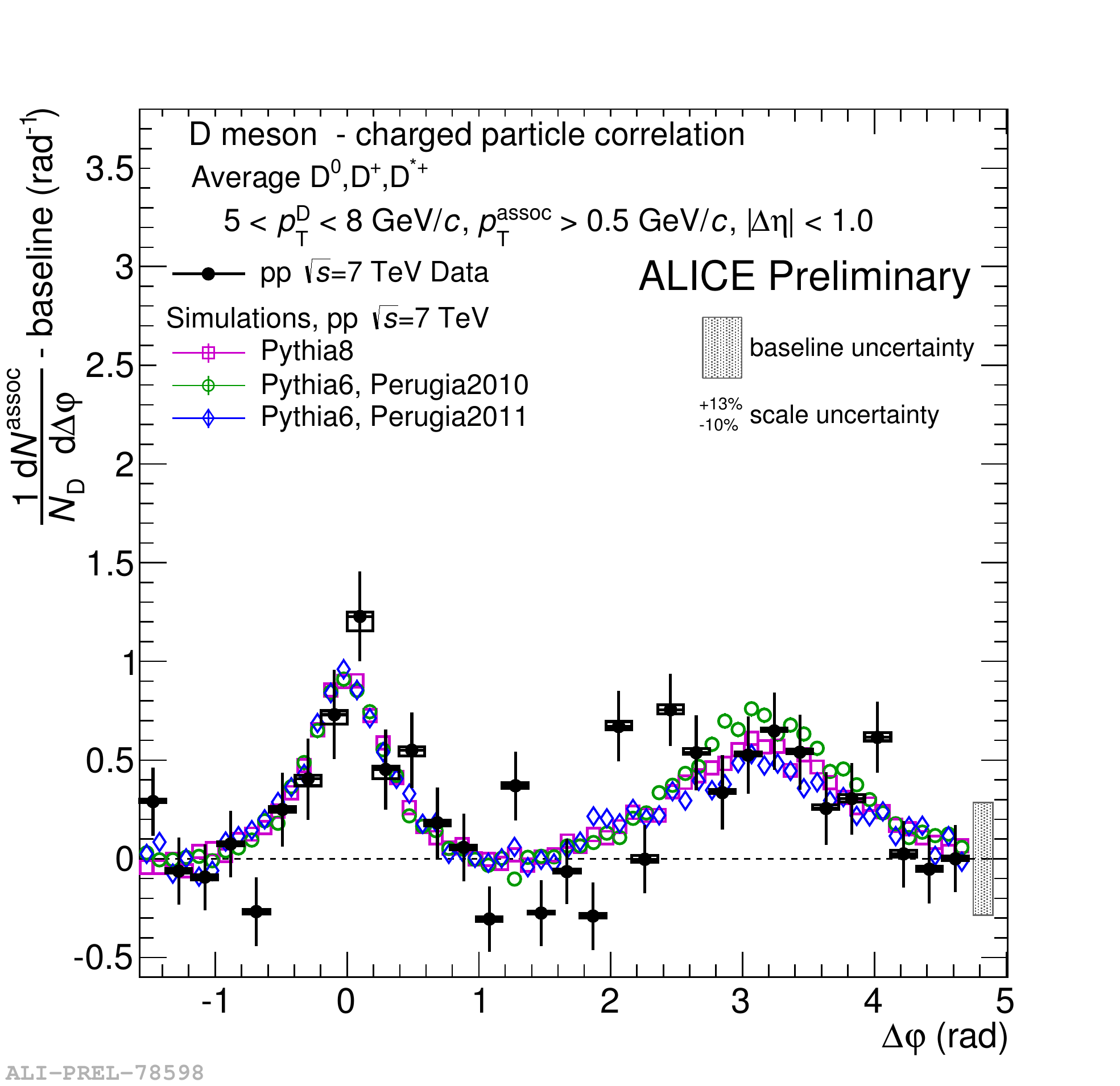}
\includegraphics[angle=0, width=6.5cm]{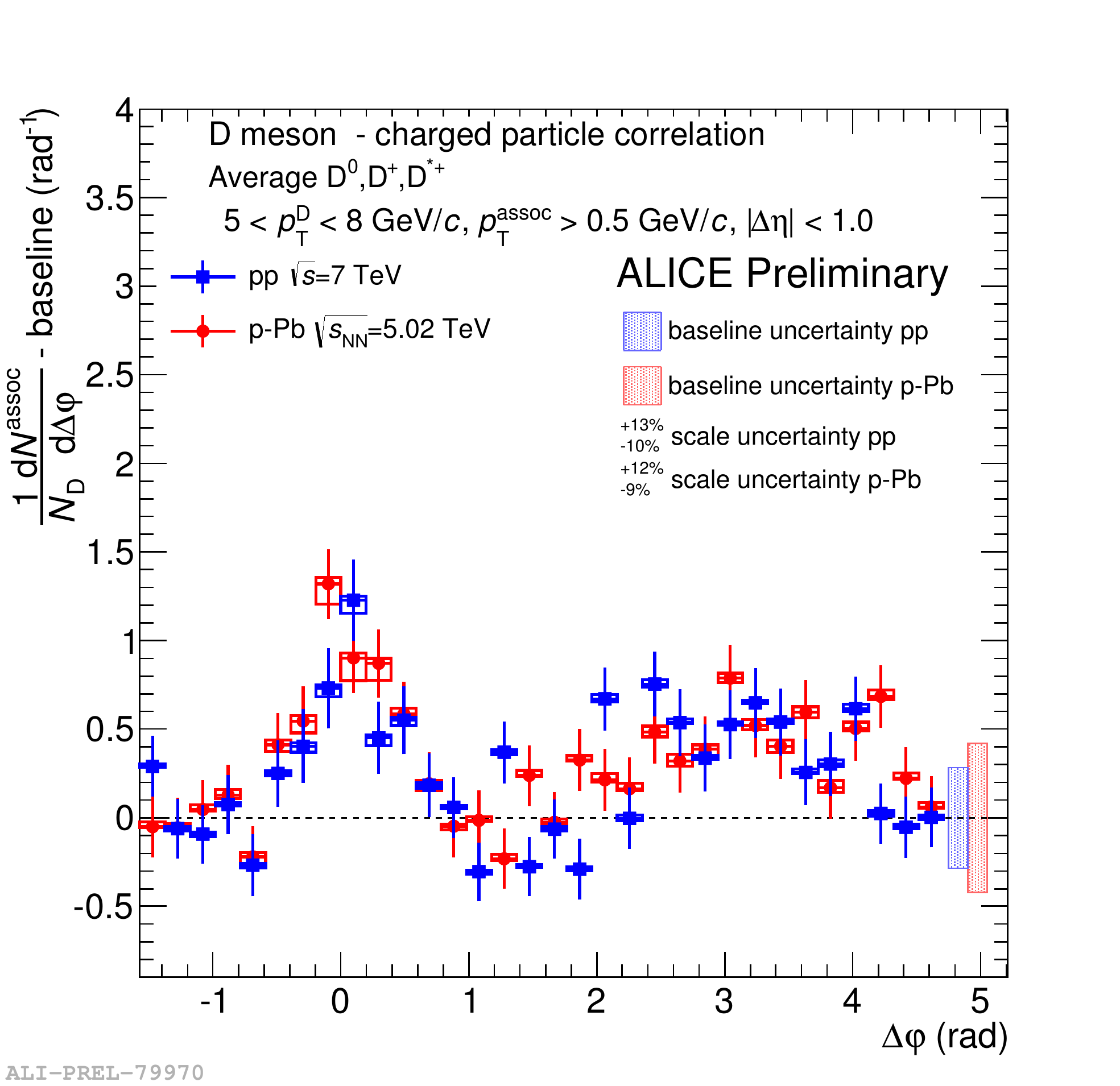}
\caption{Left:  D-hadron baseline-subtracted azimuthal correlations in pp collisions at $\sqrt s$=7 TeV, for D mesons (weighted average of D$^0$, D$^+$, D$^{*+}$) with $5<p^{\rm D}_{\rm T}<8$ GeV/$c$ and associated hadrons with $p^{\rm {assoc}}_{\rm T}>0.5$ GeV/$c$. The results are compared to PYTHIA simulations at the same energy using different tunings of the simulation parameters. Right: D-hadron azimuthal correlation distributions in pp (blue squares) and p--Pb collisions
(red circles), after baseline subtraction, for  $5<p^{\rm D}_{\rm T}<8$ GeV/$c$ and $p^{\rm {assoc}}_{\rm T}>0.5$ GeV/$c$. In both panels, the statistical and uncorrelated systematic uncertainties are shown as error bars and boxes respectively, while the correlated systematic uncertainties are indicated with labels.}
\label{fig:Dh}
\end{figure}

\begin{figure}[!htbp]
\centering
\includegraphics[angle=0, width=7.cm]{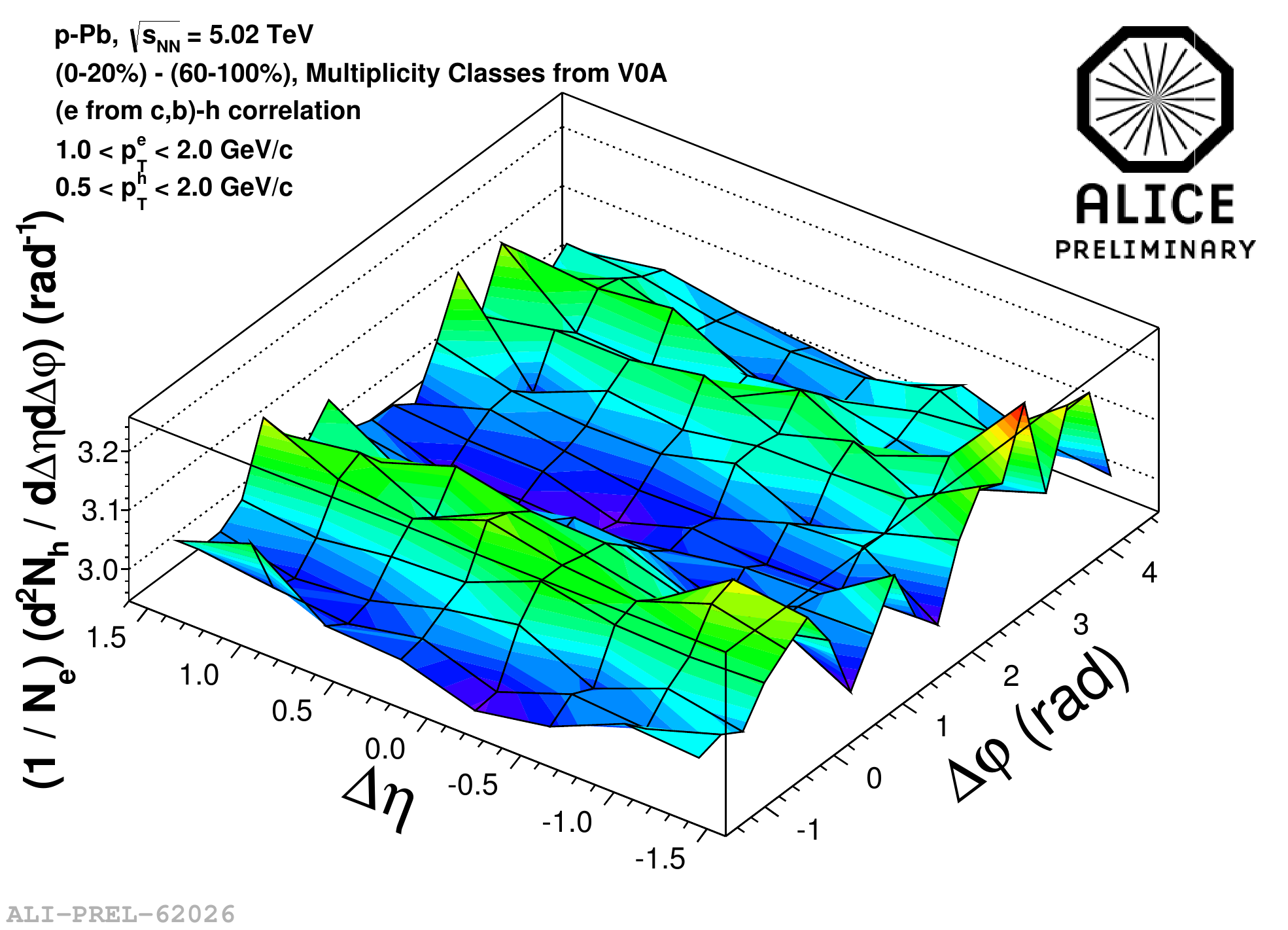}
\caption{Difference between the two-particle correlation distributions (where the trigger particle is an electron from heavy-flavour hadron decays) in high (0-20\%) and low (60-100\%) multiplicity p--Pb collisions in the ($\Delta\eta,\Delta\varphi$) space, for $1<p^e_{\rm T}<2$ GeV/$c$ and  $0.5<p^{\rm{assoc}}_{\rm T}<1$ GeV/$c$.}
\label{fig:eh}
\end{figure}

Measurements of azimuthal correlations between heavy-flavour hadrons (or their decay products) and charged hadrons are being carried out in pp collisions with ALICE to study charm quark jets and their fragmentation properties. Similar measurements are of particular interest also in p--Pb collisions to study cold nuclear matter effects on the charm jet structure and to search for long-range correlations in both near ($\Delta \varphi \sim 0$) and away ($\Delta \varphi \sim \pi$) side regions (``double ridge''), as observed with hadron-hadron correlations at LHC energies~\cite{ALICEdoubleridge,CMSdoubleridge,ATLASdoubleridge}.
The angular correlations are built by associating a reconstructed D meson or an electron coming from a heavy-flavour hadron decay (trigger particle) with the charged hadrons (associated particles). Corrections for limited detector acceptance, spatial inhomogeneities, trigger and associated particle selection efficiency, secondary track contamination are applied. 
In the case of D-hadron correlations, the fraction of D mesons coming from beauty-hadron decays was estimated with FONLL~\cite{fonll} and it ranges between 5\% and 20\% depending on the $\pt$ interval. A template distribution of angular correlations of D mesons from beauty-hadron decays and charged hadrons was obtained with PYTHIA simulations and subtracted from the measured D-hadron correlations.
Figure~\ref{fig:Dh} (left) shows the D-hadron azimuthal correlations obtained in pp collisions at $\sqrt s$=7 TeV
and compared with the results extracted from PYTHIA simulations at the same energy using different tunings of the simulation parameters. The azimuthal correlations from both data and simulations are reported after the subtraction of the baseline. An overall compatibility within the uncertainties is found between data and the PYTHIA simulations. 
The comparison to p--Pb collisions is shown in the right panel of Fig.~\ref{fig:Dh}, after baseline subtraction. Comparable results within uncertainties are obtained in pp collisions at  $\sqrt s$= 7 TeV and p-Pb collisions at $\sqrt s_{NN}$ = 5.02 TeV. The current limited statistics is preventing us from drawing quantitative conclusions on possible cold nuclear matter effects on the charm jet properties.
As observed for light-flavour hadrons, angular correlations can provide further insight into particle production mechanisms and cold nuclear matter effects. In particular, in di-hadron correlations a long-range structure was observed on both near and away sides in high-multiplicity p--Pb collisions\cite{ALICEdoubleridge,CMSdoubleridge,ATLASdoubleridge} and different mechanisms
were advocated to explain its origin, either related to initial-state effects (i.e. Color Glass Condensate~\cite{CGC1,CGC2}), or to final-state effects like collective motion explained by hydrodynamics~\cite{hydro1,hydro2}. 
It becomes therefore interesting to study whether such mechanisms are affecting also the heavy-flavour sector.
The two-particle angular correlations between heavy-flavour decay electrons and charged particles were measured in p--Pb collisions at $\sqrt s_{NN}$ = 5.02 TeV, in the high (0-20\%) and low (60-100\%) multiplicity classes. The correlated jet contribution was reduced by subtracting the ($\Delta\eta,\Delta\varphi$) correlations obtained in low-multiplicity events from those obtained in the high-multiplicity class. The subtracted ($\Delta\eta,\Delta\varphi$) correlations are shown in Fig.~\ref{fig:eh}. Two long-range structures, on both near and away sides, are seen at relatively low electron $\pt$, similarly to what was measured for di-hadron correlations.

\section{Conclusions}

In summary, the cross sections of D mesons and leptons from heavy-flavour decays measured with ALICE in pp collisions at $\sqrt s$ =7 and 2.76 TeV are well described by different perturbative QCD calculations, confirming that heavy  quarks provide a powerful test for pQCD at LHC energies. 
The ALICE heavy-flavour measurements in different rapidity ranges in p--Pb collisions are relevant to study the nuclear parton distribution functions at small x and to explore different initial-state effects at different rapidities.
The D-meson self-normalized yields exhibit an increasing trend with increasing charged-particle multiplicity in pp collisions, suggesting that MPIs affect the hard momentum scale relevant for heavy-flavour production.  A similar increase is observed in p--Pb collisions, but in this case a higher number of binary nucleon-nucleon collisions also contributes to high-multiplicity events.
A double-ridge structure was observed in two-particle correlation distribution triggered by heavy-flavour decay
electrons in high-multiplicity p--Pb collisions with ALICE. This is similar to what was observed for light-flavour hadrons, indicating that the responsible mechanisms (initial/final-state effects) might
also affect heavy quarks.

\end{document}